\documentclass[pra,aps,amssymb,footinbib]{revtex4}
\usepackage{amssymb}

\usepackage{amssymb}

\usepackage{graphicx}
\usepackage{amsmath}
\usepackage{times}
\usepackage{color}

\begin{document}

\title{Noise Correlations of Hard-core Bosons: Quantum Coherence and Symmetry Breaking }
\author{ Ana Maria Rey $^{1,2}$ \footnote{Electronic address: arey@cfa.harvard.edu} , Indubala I
 Satija $^{2,3}$ \footnote{Electronic address: isatija@physics.gmu.edu} and Charles W Clark$^{2}$ }

\address{$^{1}$Institute for Theoretical
Atomic, Molecular and Optical Physics, Harvard-Smithsonian Center of
Astrophysics, Cambridge, MA, 02138.}

\address{$^{2}$National Institute of Standards and Technology,
Gaithersburg, MD 20899}

\address{$^{3}$  Dept. of Phys., George Mason U., Fairfax, VA,
22030}

\date{\today}

\begin{abstract}
Noise correlations, such as those observable in the time of flight
images of a released cloud, are calculated for hard-core bosonic
(HCB) atoms. We find that the standard  mapping of HCB systems onto
spin-$1/2$ XY models fails in application to computation of noise
correlations due to the contribution of {\it multiply occupied
virtual states} in HCB systems. Such states do not exist in spin
models. An interesting manifestation of such states is the breaking
of particle-hole symmetry in HCB. We use noise correlations to
explore quantum coherence of strongly correlated bosons in the
fermionized regime with and without external parabolic confinement.
Our analysis points to distinctive new experimental signatures of
the Mott phase.
\end{abstract}

 \maketitle

In recent years, great experimental progress has been achieved in
the coherent control of ultra-cold gases, which has been used to
attain laboratory demonstrations of iconic model quantum many-body
systems. For example, by loading a Bose-Einstein condensate (BEC)
into a tight two-dimensional optical lattice, an array of
one-dimensional tubes has been created
\cite{Tolra,Weiss,Paredes,Moritz,Fertig}. Using this setup, recent
experiments have been able to successfully enter the Tonks-Girardeau
(TG) regime \cite{Weiss,Paredes} predicted many years ago \cite{GR}.
Furthermore, by using an additional one-dimensional optical lattice
in the direction of the tubes, a one-dimensional Mott insulator (MI)
state with unit filling has been realized experimentally
\cite{Paredes}.  In the TG limit, bosons behave like impenetrable
particles - hard-core bosons (HCB) - and their dynamics are
equivalent in many respects to those of a system of fermions. In
this paper, we calculate four-point correlation functions of HCB and
the corresponding spin-$1/2$ and fermion systems.

Experimental access to such correlation functions has recently been
achieved \cite{Greiner,Foelling} by analysis of spatial (noise)
correlations in shot-noise-limited images of expanded atomic clouds,
following the original suggestion of Altman {\it et al.}
\cite{Altman}. For example, for the case of bosons in a
three-dimensional MI state, these noise correlations have been
proven to complement the standard, first-order, characterization of
phase coherence that is expressed by the direct image of the density
of the expanded gas \cite{Altman,Foelling}.  Such correlations have
also been used to probe pair correlations of fermionic atoms
generated by molecular dissociation \cite{Greiner}.

In this paper, we develop a theoretical framework to compute noise
correlations for HCB systems confined in a one-dimensional lattice.
The Hamiltonian of this HCB system is related to that of the
spin-$1/2$ XY model, which in turn can be mapped onto the
Hamiltonian of a spinless fermion system \cite{Lieb,LM}. These three
systems have identical spectra and local observables, but their
off-diagonal correlation functions differ. Experimentally relevant
two-point correlation functions are identical for HCB and spin-$1/2$
systems, and the general formulation to calculate these correlations
was developed by Lieb and Mattis \cite{Lieb, LM}, based on Wick's
theorem. To our knowledge, explicit formulae to calculate higher
order correlations have not been worked out previously, except in
certain asymptotic limits \cite{KBI}.

Here we calculate all four-point correlation functions and use them
to compute noise correlations. One of the central results of this
paper is the discovery of important differences between noise
correlation functions of HCB and spin-$1/2$ systems. The root of
this difference is the bosonic character of  HCB  that  allows for
{\it multiple occupancy of the virtual states} (MOV) even though the
ground state is at most singly occupied. This is in contrast to
spin-$1/2$ systems whose Hilbert space excludes multiple occupancy
of all states. Here we show that the standard Lieb and Mattis
\cite{Lieb, LM} formulation for spin $1/2$ systems can still be
utilized for calculating correlations functions in HCB systems
provided the operators are written in normal order form by using
boson commutation rules, before applying any transformation.

Our study shows that MOV states are responsible for various differences between
the HCB and the XY spin-$1/2$ correlations.
One particular way to characterize this difference is the particle-hole  symmetry:
in HCB noise correlations, this symmetry is broken while it is preserved in the corresponding
spin systems.

We use the noise correlations to characterize quantum coherence in
both homogeneous lattice systems and systems subject to external
parabolic confinement, as is often the case in experiments. Previous
studies of noise correlations have been carried out in the MI limit
\cite{Altman,Foelling}; our analysis extends this understanding into
the strongly correlated (fermionized) regime. We show that, {\it
independent of the filling factor of the system}, HCB exhibit second
order coherence.  It is manifested by peaks in the noise shot images
which reflect the order induced by the lattice potential. This
suggests that second order coherence in noise correlations is a
generic attribute of correlated bosonic systems; as was also noted
in earlier studies in a different context \cite{Altman,Mathey}, such
coherence does not merely signal reduced number fluctuations. Our
detailed study involving all fillings factors highlights
distinctions between the Mott regime and the compressible regime. In
the MI limit, the intensity of the density-density correlations
between any two points separated by a reciprocal lattice vector is
constant. In the compressible phase, on the other hand, the
corresponding noise correlations depend strongly on the spatial
position in the image. Thus a regular pattern in noise correlations
could serve as a definite signature of the Mott phase. Additionally,
the peaks of noise correlations in the compressible phase are
accompanied by satellite dips immersed in a negative background.
These dips are suppressed in the insulating phase.

In Section I, we discuss various Hamiltonians related to HCBs and
describe noise correlations that can be measured in experiments
involving cold atoms in optical lattices. In Section II, we develop
the theoretical framework needed to calculate all four point
correlation functions relevant for computing noise correlations. In
view of the relationships among HCB, XY spin-$1/2$, and spinless
fermion models, we develop explicit formulas to calculate noise
correlations for all of them. In Section III, we use the formulation
of section II to study noise correlations for a spatially
homogeneous gas of HCB. We discuss their basic properties and
describe in detail how the strengths of the peaks and dips depend
upon the filling factors. Our detailed numerical studies suggest
that although the peak and dip amplitudes depend upon the size of
the system, the peak to dip  ratio  is size independent. We compare
noise correlations for HCB, spin-$1/2$ XY chains, and the
corresponding fermionic systems, and show that the spin noise
correlations preserve the particle-hole symmetry of the Hamiltonian,
but that this symmetry is broken in HCBs. In Section IV, we compute
the noise correlations for experimentally relevant parameters and
compare results for Mott vs. compressible phases. Section V
summarizes our results.

\section{ Introduction and Formulation }
\subsection{Bose-Hubbard Hamiltonian and Observable}
The Bose-Hubbard Hamiltonian describes bosons in optical lattices
when the lattice is loaded in such a way that only the lowest
vibrational level of each lattice site is occupied and tunneling
occurs only between nearest-neighbor sites \cite{Jaksch, Sachdev}:
\begin{equation}
\hat{H}= -J\sum_{\langle i,j\rangle
}\hat{a}_i^{\dagger}\hat{a}_{j}+\frac{U}{2}\sum_{j}\hat{n}_j(\hat{n}_j-1)
+ \sum_{j} V_j \hat{n}_j
\end{equation}
\noindent Here $\hat{a}_j$ is the bosonic annihilation operator of a
particle at site $j$, $\hat{n}_j=\hat{a}_j^{\dagger}\hat{a}_{j}$,
and the sum $\langle i,j\rangle$ is over nearest neighbors. The
hopping parameter $J$, and the on-site interaction energy  $U$ are
functions of the lattice depth. $V_j$ represents any other external
potential such as a parabolic confinement or on-site disorder.

In a typical experiment, atoms are released by turning off the
external potentials at time $t=0$. The atomic cloud expands, and is
photographed after it enters the ballistic regime. Assuming that the
atoms are noninteracting from the time of release, properties of the
initial state can be inferred from the spatial images. As explained
in detail in Ref. \cite{Roth,Altman},  the density distribution
after the release, $\langle \hat{n}[x]\rangle=\langle
\hat{\psi}(x,t)^{\dagger}\hat{\psi}(x,t)\rangle$,  with
$\hat{\psi}(x,t)$ the bosonic field operator, can be written in
terms of the first band annihilation and destruction operators at
lattice sites as   $\sum_{i,j} w^*(x-ia,t)
w(x-ja,t)\langle\hat{a}_i^{\dagger}\hat{a}_{j} \rangle$ where
$w(x,t)$ is the free evolution of a Wannier function centered around
the origin and $a$ is the lattice spacing. Assuming the Wannier
functions can be described by Gaussians with initial width
$\sigma_o/2$, after time of flight $t$, the spatial density at
position $x$, $\langle \hat{n}[x]\rangle$, can be written as

\begin{eqnarray}
\langle \hat{n}[x]\rangle &=& w^2(x,t) \sum_{n,m} e ^{iP(x,t) a[n-m]
}
\langle \hat{a}_n^\dagger \hat{a}_m\rangle \propto \langle \hat{n}_Q\rangle,\\
\langle \hat{n}_Q\rangle& \equiv& \frac{1}{L} \sum_{n,m} e ^{iQ
a(n-m) } \langle \hat{a}_n^\dagger \hat{a}_m\rangle,
\end{eqnarray}where   $w^2(x,t)= \sqrt{\frac{2}{ \pi \sigma ^2}}e^{-2x^2
/\sigma^2}$ and $ \sigma=\hbar t/(M\sigma_o) $. The wave-vector
$P(x,t)~=~ M x/(\hbar t)$ defines a correspondence between the
position in the expanding image, $x$, and the lattice wave vector,
$Q$. Here $M$ is the particle mass and $L$ is the number of lattice
sites.

Density-density correlations in the expanding cloud,
$\mathcal{G}[x,x'] $,   can also be linked to initial correlations in the
lattice. For the  evaluation of  $\mathcal{G}[x,x'] $  it is
important to keep in mind that only normal ordered expectation
values  can be safely replaced  by its projection into the lowest
band.  The second order correlation $\mathcal{G}[x,x'] $ is given by \cite{Altman}:

\begin{eqnarray}
&&\mathcal{G}[x,x'] = \nonumber \\
&&(w(x)w(x'))^2 \sum_{n,m,l,j} e ^{i P(x,t) a[n-m] } e ^{i
P(x',t)a[l-j] } \langle \hat{a}_n^\dagger \hat{a}_l^\dagger
\hat{a}_j \hat{a}_m\rangle\nonumber
\\&& +w^2(x)
\delta(x-x')\langle\hat{n}[x]\rangle-\langle\hat{n}[x]\rangle\langle\hat{n}[x']\rangle
\label{gxx}
\end{eqnarray} where $\delta(x-x')=L \delta_{x,x'}$ is a delta function which arises from normal ordering in the continuum.
Eq. (\ref{gxx}) can be mapped to quasi-momentum correlations in the
lattice, $\mathcal{G}[x,x'] \rightarrow\Delta(Q,Q')$, as:

\begin{eqnarray}
 \Delta(Q,Q')&\equiv&
\langle\hat{n}_Q \hat{n}_{Q'}\rangle
-\langle\hat{n}_Q\rangle\langle\hat{n}_{Q'}\rangle \nonumber \\
&-&\langle\hat{n}_{Q}\rangle(\delta_{Q,Q'+ nK}-
\delta_{Q,Q'}),\label{nnoise}
\end{eqnarray}

\noindent where $K=2\pi/a$ is the reciprocal lattice vector and $n$
is an integer. The first  term  in the second line of Eq.
(\ref{nnoise}) arises due to the bosonic commutation relations of
the operators and the identity  $1/L\sum_{l=1}^L e^{i 2 \pi m nl/L
}= \delta_{m,nL}$. The quantity $\Delta(Q,Q') \equiv \Delta_{qq'}$,
that corresponds to density-density correlations in the expanding
cloud of atoms, will be referred to as the {\it noise correlations}.
The integer $q$ characterizes the discrete values of the
quasimomentum $Q=\frac{2\pi}{La}q$ and this convention is going to
be used hereafter.

\subsection{Hard Core Bosons (HCB)}

In the strongly correlated  regime, Eq. (1) can be replaced by the
HCB Hamiltonian \cite{Sachdev},
\begin{equation}
\hat{H}^{(HCB)}=-J\sum_j(\hat{b}_j^{\dagger}\hat{b}_{j+1}
+\hat{b}_{j+1}^{\dagger}\hat{b}_{j})+\sum_{j} V_j \hat{n}_j
\label{HCB}
\end{equation}
Here $\hat{b}_j$ is the annihilation operator at the lattice  site
$j$ which satisfies $[\hat{b}_{i\neq j},\hat{b}_j^{\dagger}]=0$, and
the on-site conditions ${\hat{b}_j}^2={\hat{b}_j^{\dagger}}{}^2=0$,
which suppress multiple occupancy of lattice sites. The same
relations are fulfilled by spin-$1/2$ raising and lowering
operators, $\hat{\sigma}_j^\pm$, and this is the reason why the HCB
Hamiltonian is mapped to the spin $1/2$ -XY Hamiltonian,

\begin{eqnarray}
\hat{H}^{(\sigma)}=-2J\sum_j(\sigma_j^x\sigma_{j+1}^x+\sigma_j^y
\sigma_{j+1}^y)+V_j\sum_j  \frac{\sigma_j^z+1}{2} \label{spin}
\end{eqnarray}
Here $\sigma_l$ are Pauli matrices that satisfy anti-commutation
relation at same sites. The parameter $J$ describes the exchange
interaction between the XY-spins. A fully polarized state with all
spins up
 corresponds to the Mott insulator state.
We would like to emphasize that the mapping between the HCB and the
spin Hamiltonian is a partial mapping as the two differ in the
onsite commutation relation.

For spin-$1/2$ systems, correlation functions are in general
calculated by using the Jordan-Wigner transformation (JWT)
\cite{Lieb,Sachdev} which maps spin operators into fermionic
operators $\hat{\sigma}_j^{+}~=~\hat{c}_j^{\dagger}\exp\left[-\pi  i
\sum_{n=1}^{j-1} \hat{c}_n^{\dagger}\hat{c}_n \right]$. Here
$\hat{c}_j$ are fermion operators that obey anticommutation
relations. This transformation maps the spin Hamiltonian into a
non-interacting fermionic Hamiltonian,
\begin{eqnarray}
H^{(F)}=-J\sum_j(\hat{c}_j^{\dagger}\hat{c}_{j+1}
+\hat{c}_{j+1}^{\dagger}\hat{c}_{j})+\sum_{j} V_j \hat{n}_j.
\end{eqnarray}

Two-point HCB correlations have been calculated by invoking the
HCB$\rightarrow$ spin-$1/2$ $\rightarrow$ fermion correspondence
\cite{Rigol}. As mentioned above, in contrast to the spin operators
which  obey fermionic commutation relations  at the same site, HCB
operators at equal sites obey bosonic commutation relations.
Correlation functions such as $\langle \phi
|\hat{O}\hat{O}^{\dagger}|\phi \rangle$ (where we have suppressed
the lattice index for convenience ) are zero if $\hat{O}$ are spin
lowering operators and non-zero if $\hat{O}$ are  HCB field
operators. The non-vanishing of these correlations can be traced to
the fact that the HCB may have virtual states that have
 multiple occupancy: Recall that the HCB describes the
$U\to \infty$ limit of the Bose-Hubbard Hamiltonian, and therefore,
the limiting value of the Bose-Hubbard correlation functions must
correspond to those of HCB. In general  for  bosons one can write  $
|\phi\rangle=A|0\rangle+B|1\rangle+\epsilon|2\rangle+\dots$ where
$\epsilon\sim 1/U$ and $A$, $B$ are constants.  Therefore, $\langle
\phi| \hat{a}\hat{a}^{\dagger}|\phi\rangle= A^2+2 B^2+3
\epsilon^2+\dots$. Taking the $U\to \infty$ limit one gets $\langle
\phi| \hat{a}\hat{a}^{\dagger}|\phi\rangle\rightarrow A^2+2B^2>0$.
On the other hand this correlation is always zero for the spin
systems.

The differences in the on-site commutation rules for bosons and
spins have no effect on the calculations of local observables such
as the density and the momentum distribution. However, correlation
functions that involve a pair of operators $\hat{b}_j
\hat{b}_j^\dagger$, are affected by  MOV states. In such cases, a
direct application of the JWT is not correct. A simple recipe to
take into account the MOV problem  is to replace $\hat{b}_j
\hat{b}_j^\dagger$ by   $1+ \hat{b}_j^\dagger \hat{b}_j$ before
applying the JWT. We will refer to this recipe as the MOV rule. The
validity of the  MOV  rule was checked by comparing various
correlation functions obtained using the above recipe with those
obtained by diagonalizing a full Bose-Hubbard Hamiltonian with a
large $U$ value (see below).

\section{ Four-Point Correlations}

We now describe our calculations of four-point correlations. In view
of the  rather complex nature of the calculations, we will omit
various technical details but focus on describing the final formulas
and notations so that our results can be used for related future
work. In addition to HCB, we will also discuss the related fermion
and  spin systems.

For ideal fermions, four point correlation functions can be
calculated using Wick's theorem directly. For example, the
four-point correlations relevant for noise correlation are given by:

\begin{equation}
 \langle
\hat{c}_n^\dagger \hat{c}_m \hat{c}_l^\dagger \hat{c}_j\rangle =
-\delta_{lm}g_{nj}+g_{lj}g_{nm}-g_{lm}g_{nj},
\end{equation}
where $g_{lm}$ are free-fermionic propagators:
\begin{equation}
g_{lm}=\sum_{s=0}^{N-1} \psi_l^{*(s)} \psi_m^{(s)},
\end{equation} with $N$ being the total number of atoms and $ \psi_l^{(s)}$  the
$s^{th}$ eigenfunctions of the single-particle Hamiltonian
$-J(\psi_{l+1}^{(s)}+\psi_{l-1}^{(s)})+V_l \psi_{l}^{(s)}=E^{(s)}
\psi_{l}^{(s)}$.

For HCB and spin systems, the calculation of four-point functions is
 more complex.
Specifically, our calculations involve the following steps. (i)
rearrange the operators so that the site index is ordered (this is
only relevant for the case when three or more site indices are
different); (ii) write operators in normal order form using bosonic
rules to take into account the MOV; (iii) use the JWT accordingly to
the prescription of Lieb and Mattis; (iv) use Wick's theorem to
write higher order correlations in terms of two-point correlations.

To present our results we denote the creation and the annihilation
operators by $\hat{b^{\alpha}}$, where $\alpha =+1(-1)$ for
annihilation (creation) operators, respectively, and label the "site
ordered" four-point correlation function by
$\chi^{\alpha\beta\gamma\delta}_{abcd}$

\begin{equation}
 \langle
\hat{b}_n^\dagger \hat{b}_m \hat{b}_l^\dagger \hat{b}_j\rangle
\mapsto \langle \hat{b}_{a}^{(\alpha)} \hat{b}_{b}^{(\beta)}
\hat{b}_{c}^{(\gamma)} \hat{b}_{d}^{(\delta)}\rangle \equiv
\chi^{\alpha\beta\gamma\delta}_{abcd},
\end{equation}

In this equation it is implicit that $a\leq b \leq c \leq d$. This
order is implied in all the expressions that follow. We define
$G_{ij}\equiv 2 g_{ij} -\delta_{i,j}$ and $B_{ij}\equiv \langle
\hat{b}_{i}^{(\dagger)} \hat{b}_{j}\rangle$. The latter can be
calculated in terms of $G_{ij}$ \cite{LM}. We introduce four
matrices ${\bf M},{\bf S},{\bf X}$ and ${\bf Y}$  in terms of which
our results for the correlation functions can be written as

\begin{eqnarray}
&&\chi^{\alpha\beta\gamma\delta}_{abbd}=\frac{ \beta
\gamma-1}{2}\left[\frac{(-1)^{s \eta_\beta}}{4}  |{\bf M}(a,b,d
)|-\left(\frac{1}{2}+(1-s) \delta_{\beta,-1}\right) B_{ad} \right],\\
&& \chi^{\alpha\beta\gamma\delta}_{aacd}=\frac{1-
\alpha\beta}{2}\left[\frac{(-1)^{s \eta_\alpha} }{4}   |{\bf
S}(a,c,d)|+\left(\frac{1}{2}+(1-s) \delta_{\alpha,-1}\right)
B_{cd}\right],\\
&&\chi^{\alpha\beta\gamma\delta}_{abcc}=\frac{1- \gamma
\delta}{2}\left[\frac{1}{4} (-1)^{s \eta_\gamma} |{\bf
S}(c,a,b)|+\left(\frac{1}{2}+(1-s)\delta_{\gamma,-1}\right)
B_{ab}\right],\label{noisec}\\
&&\chi^{\alpha \beta \gamma \delta}_{abcd}=(-1)^
{b+d-c-a}\left[\frac{2- \gamma\delta- \alpha\beta }{16} |{\bf
X}(a,b,c,d )|+
\frac{\beta}{4}(\delta_{\gamma,-1}-\delta_{\alpha,-1}) |{\bf
Y}(a,b,c,d )|\right].
\end{eqnarray}
\begin{eqnarray}
{\bf M}&=&\left(\begin{array}{ccccc}
  G_{aa+1} & G_{ab-1} & G_{ab+1} & .. & G_{ac} \\
  \vdots &  & & & \vdots \\
  G_{b-1a+1} & G_{b-1b-1} & G_{b-1b+1}& .. & G_{b-1c} \\
  G_{b+1a+1} & G_{b+1b-1} & G_{b+1b+1}& .. & G_{b+1c} \\
  \vdots &  &  & & \vdots \\
  G_{c-1a+1} & G_{c-1b-1} & G_{c-1b+1}& .. &
  G_{c-1c}\end{array}\right), \quad
 \\ {\bf X}&=&\left(\begin{array}{ccccccc}
  G_{ba} &  .. & G_{bb-1} & G_{bc+1} & .. & G_{bd-1}& G_{bc}\\
  G_{a+1a}& .. & G_{a+1b-1}& G_{a+1c+1} & .. & G_{a+1d-1}&
  G_{a+1c}\\
    \vdots &  &  & & &\vdots \\
    G_{b-1a}& .. & G_{b-1b-1}& G_{b-1c+1} & .. & G_{b-1d-1}&
  G_{b-1c}\\
  G_{c+1a}& .. & G_{c+1b-1}& G_{c+1c+1} & .. & G_{c+1d-1}&
  G_{c+1c}\\
   \vdots &  &  & & &\vdots \\
  G_{da}& .. & G_{db-1}& G_{dc+1} & ..& G_{dd-1}&
  G_{dc}
\end{array}\right),
\end{eqnarray}

\begin{eqnarray}
{\bf S}&=&\left(\begin{array}{cccc}
  G_{aa} & G_{ab+1} & .. & G_{ac} \\
  G_{ba} & G_{bb+1} & .. & G_{bc} \\
  \vdots &  &  & \vdots \\
  G_{c-1a} & G_{c-1b+1} & .. & G_{c-1c}
\end{array}\right),\\ {\bf Y}&=&\left(\begin{array}{cccccccc}
  G_{a+1a} & G_{a+1b} & G_{a+1a+1} & .. & G_{a+1b-1}& G_{a+1c+1} & ..& G_{a+1d-1}\\
 \vdots &  &  & & && &\vdots \\
  G_{b-1a} & G_{b-1b} & G_{b-1a+1} & .. & G_{b-1b-1}& G_{b-1c+1} & ..&G_{b-1d-1}\\
 G_{c+1a} & G_{c+1b} & G_{c+1a+1} & .. & G_{c+1b-1}& G_{c+1c+1} &..& G_{c+1d-1}\\
  \vdots &  &  & & & & &\vdots \\
G_{d-1a} & G_{d-1b} & G_{d-1a+1} & .. &
G_{d-1b-1}& G_{d-1c+1} & ..&G_{d-1d-1}\\
G_{ca} & G_{cb} & G_{ca+1} & .. &
G_{cb-1}& G_{cc+1} & ..&G_{cd-1}\\
G_{da} & G_{db} & G_{da+1} & .. & G_{db-1}& G_{dc+1} &..& G_{dd-1}
\end{array}\right).
\end{eqnarray}

\noindent Here $s$ is equal to  0 for HCB and $1$ for spins and
$\eta_\beta=\delta_{\beta,-1}$.  In the matrices dots indicate
continuous variation of the indices. In the absence of dots, the
indices are explicitly written and they may not change continuously.

When three, four or two pairs of indices are equal the calculation
of the  four-point functions is simple and no determinant is
actually  required. The non-vanishing correlation functions of this
type are given by:

\begin{eqnarray}  \langle \hat{b}_n^\dagger\hat{b}_n
\hat{b}_n^\dagger \hat{b}_n\rangle&=&g_{nn},\\
\langle\hat{b}_n^\dagger \hat{b}_n\hat{b}_m^\dagger
\hat{b}_m\rangle&=&g_{nn} g_{mm}- g_{nm}^2,\\  \langle
\hat{b}_n^\dagger \hat{b}_m\hat{b}_m^\dagger \hat{b}_n\rangle&=&
(-1)^s(g_{nn} g_{mm}- g_{nm}^2)+ g_{nn},\\ \langle\hat{b}_n^\dagger
\hat{b}_n \hat{b}_n^\dagger \hat{b}_m\rangle&=&\langle
\hat{b}_m^\dagger \hat{b}_n \hat{b}_n^\dagger
\hat{b}_n\rangle=B_{nm}.
\end{eqnarray} with $n\neq m$ and  $B_{ij}\equiv \langle
\hat{b}_{i}^{(\dagger)} \hat{b}_{j}\rangle$. The latter can be
calculated in terms of $G_{ij}$ \cite{LM}.

\section{ Noise Correlations for Homogeneous Systems }

We now analyze various characteristics of the noise correlations
that are calculated using the formulation discussed above. We
consider particles in a periodic lattice in the absence of any
external potential ($V_i=0$). Here we compare and contrast HCB, XY
spin-$1/2$ and the corresponding fermion system. This comparison
illustrates the importance of MOV and the quantum statistics of the particles
in noise correlations.
Our study includes all
filling factors $\nu$, $0 \le \nu \le 1$. Here $\nu=N/L$ where
 $L$ is the number of
lattice sites and $N$ is the total number of particles. We thus
study the Mott regime ($\nu=1$) as well as the non-Mott regime with
$ 1 < \nu < 1$.

For  the HCB and the spin systems the T\"{o}plitz-like determinants
involved in the evaluation of noise correlations make the
calculations complicated.  The only simple case that can be studied analytically
is  when the
system is a MI ( $\nu=1$). In this case,
$g_{ji}=
\delta_{i,j} $ and the determinants involved in the calculation become trivial.
 The noise correlations are given by
\begin{eqnarray}
\Delta^{HCB}(Q_1,Q_2)|_{\nu=1}=\delta_{Q_1,Q_2}+
\left(\delta_{Q_1,Q_2+ j K}-\frac{2}{L}\right) \label{mott},
\end{eqnarray} where $j$ is an integer. Eq. (\ref{mott}) shows the characteristic bosonic peaks at every
 reciprocal lattice vectors, which reflect the
 order induced by the periodicity of the lattice.
The peaks in HCB correlations reflect the bunching of the bosons due
to Bose-Einstein statistics. It should be noted that the noise
correlations at $Q_1=Q_2$ (autocorrelations) are different from
those at $Q_1=Q_2+Kj$, $(j\neq 0)$. For a unit filled MI in
particular, the former is two times larger than the latter
,$\Delta(Q_1,Q_1)=2 \Delta(Q_1,Q_1+K j)$.

For spins,  $\nu=1$ corresponds to a fully polarized system. In this
case, the local correlations dominate and the spins  behave  like
fermions. The fermion correlations are given by
\begin{eqnarray}
\Delta^F(Q_1,Q_2)&=&-n_{Q_1}^2 \delta_{Q_1-Q_2,j K} \quad \quad
j\neq 0, \label{fermino}
\end{eqnarray}where $n_{Q_1}=1 $ for $[Q_1]_K< Q_F$ and zero otherwise. Here
$[]_K$ means modulo reciprocal lattice vectors,  $j$ is an integer
different from zero and  $Q_F$ is the Fermi wave-vector
$Q_F~=\frac{2\pi\parallel~\nu/2~\parallel}{a} $(where $\parallel
x\parallel $ denotes the integer part). The fermionic noise
interference pattern shows negative interference dips at every
reciprocal lattice vector (except at $Q_1=Q_2$ where
$\Delta(Q_1,Q_1) = 0$) signaling the underlying anti-bunching due to
Fermi-Dirac statistics.


From Eqs. (\ref{mott}) and  (\ref{fermino})  one arrives at the
conclusion that in the Mott regime ($\nu=1$), HCB noise correlations
with central peak height equal to $2$ differ from the corresponding
spin correlations which show  no central peak. Thus this simple
limit brings out the importance of MOV in HCB.

\begin{figure}[htbp]
\begin{center}
\includegraphics[width=3.5in,height=3.5in]{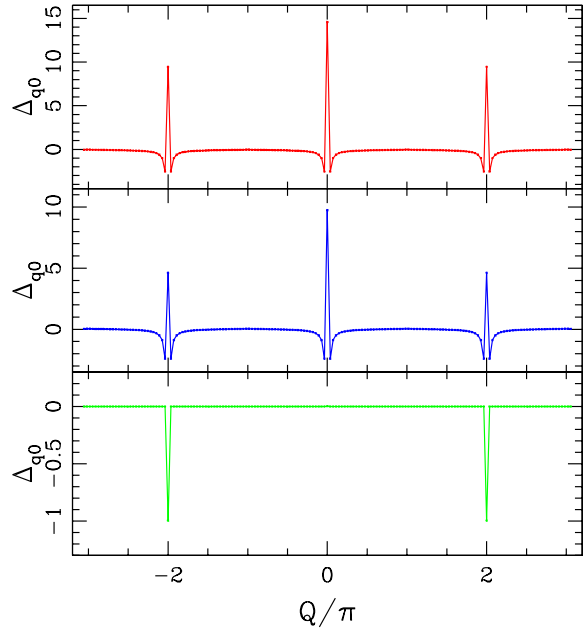} \leavevmode
\end{center}
\caption{ Noise correlations at  $Q_2=0$ for different values of
$Q=2\pi q/(La)$ (In the plot the x-axis is in units of $1/a$). The
upper, middle and lower panels respectively correspond to HCB, spins
and fermions (note the different scale used in the bottom panel). We
use $L=55$ and $N=27$.} \label{fig1}
\end{figure}

For non-integer filling factors  $0< \nu <1$, there is no simple
analytic expression for the noise correlations. Therefore, all four
point functions and the noise correlations are computed numerically.
Our detailed investigation included lattices of various sizes and we
studied both fixed end  as well as periodic boundary conditions.
Fig.~\ref{fig1} illustrates the quantum coherence of HCB, spin and
fermions as characterized by  $\Delta_{q0}$. Our calculations show
that for HCB as well as for spin systems noise correlations  exhibit
characteristic peaks at every reciprocal lattice vector for all
filling factors. These  peaks reflect  the  underlying  order
induced by the periodic lattice. The heights of the peak vary with
the filling factor (see Fig.~2). The peaks are a manifestation of
the boson nature of the atoms and are induced by interactions as
they disappear  in the non-interacting regime where all of the atoms
are Bose-condensed. For fractional values of $\nu$, the peaks are
accompanied  by a small satellite dip at $Q =2\pi/(La) + j K$
immersed in a negative background. In other words, the dips occur
only in the compressible phase. These dips are present only for
non-integer filling factor and are a manifestation of the quasi-long
range coherence of the system.

It turns out that Bogoliubov theory \cite{Bog} is helpful in gaining
insight into the
the dips and the negative background in the HCB noise correlations.
The theory describes the weakly interacting regime and in principle
can not be used to quantitatively describe the HCB gas. However,
some aspects of the theory survive in HCB limit and suggest a
possible mechanism for the dips and the negative background that
characterize the compressible regime. The Bogoliubov approximation
predicts the negative background $\Delta(0,Q)<0$ and the satellite
dip at low quasi-momenta (phonon regime). In the thermodynamic limit
$\Delta(0,Q)$ diverges as $-\frac{ \nu U}{4J Q^2}$. For finite
lattices  the height of the satellite dip does not diverge but
scales like $-\frac{ \nu U}{4J }\frac{L^2}{4 \pi^2}$. Moreover in
the Bogoliubov approximation the central peak scales like $\frac{
\nu U}{4J}\frac{ L}{2\pi} \cot[\pi/L]$ so the peak to dip ratio
remains finite (see also \cite{student}). As we discuss below, the
insensitivity of this ratio to the lattice size is also seen in HCB
case.

The negative background, $\Delta(0,Q) <0$, and the satellite dip at
low quasi-momenta can also be qualitatively understood as follows:
$\Delta(0,Q)$ can be written as $\Delta(0,Q)=\langle e| e\rangle-
N_0 n_Q$ where $N_0$ is the number of condensate atoms, $n_Q$ is the
number of atoms with quasi-momenta $\hbar Q$ (the population of
which comprises the quantum depletion)  and $\langle e| e\rangle$ is
the amplitude of the state $|e\rangle=|\hat{a}_Q
\hat{a_0}|g\rangle$, with $|g\rangle$ the many-body ground state.
$|e\rangle$  represents the process of removing a particle from the
condensate and one from the state with quasi-momentum $\hbar Q$. In
the absence of interactions all of the atoms are Bose condensed,
correlations can be neglected and $\langle e| e \rangle= N_0 n_Q=0$.
On the other hand, interactions dramatically change  the behavior of
the Bose gas. Collisions between zero-quasimomentum atoms admix into
the condensate pairs of atoms at quasimomenta $\pm \hbar Q$. As a
result pair excitations in the condensate become correlated so as to
minimize the total energy of the system giving rise to the negative
sign in the noise correlations. For high quasi-momenta the
interference plays a minor role but in the phonon regime $n_{-Q}$
diverges  and the anti-correlations are maximal.

As mentioned above for  HCB bosons Bogoliubov theory is not
qualitatively valid but it still gives a useful physical picture  as
the HCB ground state also exhibits off-diagonal quasi-long-range
order that manifests in the power law decay of the density matrix
and the macroscopic occupation of the zero quasi-momentum state
(quasi-condensate) which scales like $\sqrt{N}$ \cite{Rigol} (see
Fig. 5).

In  Fig.~2 we study the dependence of the peak and dip height with
the filling factor for HCB and the correspondent spin system. As
illustrated there the heights ($\Delta_{00}$) and the dips
($|\Delta_{10}|$) of the peaks increase with the filling factor up
to a maximum value and then begin to decrease. At $\nu=1$, for HCB,
the height approaches $2$ while the dip vanishes
($\Delta(Q,0)\rightarrow -2/L$). However, for the spin model, the
corresponding value is $0$ for both peak and dip.
\begin{figure}[htbp]

\begin{center}
\leavevmode
\includegraphics[width=3.5 in,height=3.5 in]{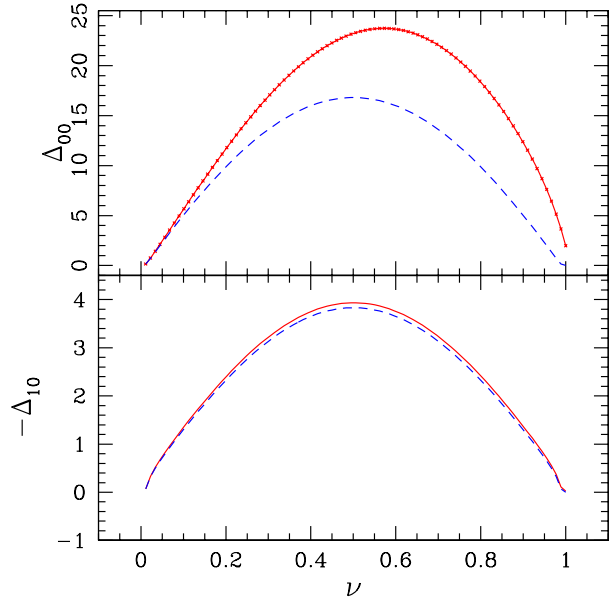}
\end{center} \caption{  The figure shows the central peak at occurs
at $q=q'=0$(top) and the  the corresponding satellite dip intensity
at $q=1$ and $q'=0$ as we vary the filling factor. Red and blue
respectively correspond to HCB and spin systems. Here $L=89$ and we
use free boundary conditions.} \label{fig2}
\end{figure}

 The central  peak attains its  maximum value at
$\nu \approx 0.6$ and the curve describing the variation of its
height with the filling factor  lacks  reflection symmetry about
$\nu=1/2$. These results were confirmed for various lattice sizes as
well as for periodic boundary conditions. The asymmetric behavior
implies that the noise correlations for HCB do not preserve particle
hole symmetry. In contrast, the satellite dip appears to preserve
it. The asymmetry in HCB correlations  has its roots in the MOV
states as the spin correlations (which do not have MOV) preserves
the particle-hole symmetry.  In fact,  the spin Hamiltonian, Eq.
(\ref{spin}) with the additional constraints
 $[\hat{\sigma}^-_{i\neq j},\hat{\sigma}_j^{+}]=0$, and
 $[\hat{\sigma}^-_{i},\hat{\sigma}_i^{+}]_+=1$,
 is particle-hole symmetric under the transformation
$\hat{\sigma}^-_i=\hat{h}_i^{\dagger}$, $\hat{\sigma}^+_i=\hat{h}_i$.
 The operators $\hat{h}_i^{\dagger}$ and $\hat{h}_i$ are the creation and
 annihilation operator for holes. On the other hand, if
 $[\hat{\sigma}^-_{i},\hat{\sigma}_i^{+}]_+=1$ is replaced by
 $[\hat{b}_{i},\hat{b}_i^{\dagger}]=1$, as is the case for HCB,
 the  particle-hole symmetry  is no longer preserved under this
 transformation (which in terms of HCB operators reads as $\hat{b}_i=\hat{h}_i^{\dagger}$,
$\hat{b}^\dagger_i=\hat{h}_i$).

It is important to mention that, in general, different observables
do not have to preserve the symmetries of the Hamiltonian.
 For example, the  momentum distribution  for
 spins has an explicit dependence on the density  and is not
 completely symmetric around  $\nu=1/2$ \cite{Rigol}. Nevertheless  as shown in Fig. 2,
noise-correlations do preserve the particle hole-symmetry of the
Hamiltonian.


Our detailed numerics shows that  the heights of the peaks and the
dips depend upon the lattice size and are expected to diverge in the
thermodynamic limit. However, our simulations also suggest that the
ratio of height to dip is size independent (behavior also observed
in the Bogoliubov calculations) for various filling factors. In
particular  for HCB the ratio near the maxima of the central peak is
found to approach $2 \pi$ (See Fig.~3). The scale invariance of this
ratio makes it a more desirable physical quantity that can be
compared in different experiments.

The striking differences between the spins and the HCB  noise
correlations not only highlight the importance of including  MOV
states but also explicitly show  that a direct  mapping  between the
two systems is not valid for all observables. We checked the
validity of MOV rule by calculating  noise correlations using a
direct diagonalization  of  the Bose-Hubbard Hamiltonian for small
size systems. Fig.~4 illustrates the variation of height of the
central peak and the dip as a function of the interaction $U$. As
$U$ increases noise correlations reach the asymptotic value that was
found to be in excellent agreement with the value calculated  from
the HCB model including MOV states. We also calculate the noise
correlations for the spin model, to emphasize the fact that MOV
corrections can be significant. Even for a small size system, Fig.4
clearly shows broken symmetry for HCB while the symmetry is
preserved for the spin system. In Fig. 4  the central peak is
found to be associated with a dip for all  values of $U$ in
consistency with the Bogoliubov calculations.
\begin{figure}[htbp]
\begin{center}
\includegraphics[width=3.5 in,height=3.5 in]{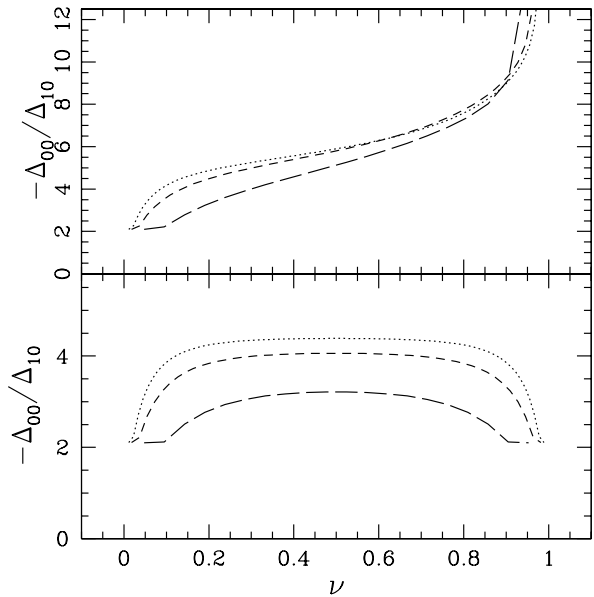}
\leavevmode \end{center}
 \caption{  The upper and the lower panels
respectively show the ratio of height to dip for the HCB and spin
systems. The three curves in each panel respectively correspond to
$L=89$ (dotted), $L=55$ (short dashed), $L=21$ (long dashed) lines.}
\label{fig3}
\end{figure}

\begin{figure}[htbp]
\begin{center}
\includegraphics[width=3. in,height=4 in]{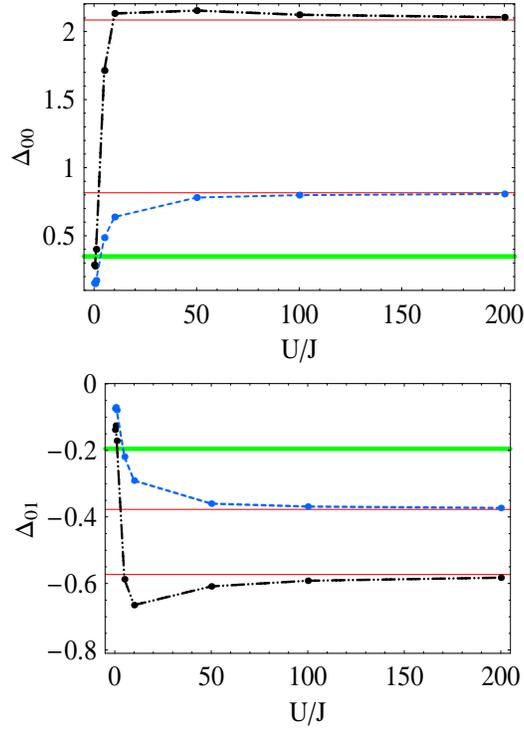}
\leavevmode
\end{center} \caption{ The figure shows the peak (upper) and the
dip(lower) intensities for $L=6$ and $N$ equal to $2$(dashed blue
line) and $4$ (dotted-dashed black) as the interaction $U$ varies.
The solid horizontal red lines show the corresponding result for HCB
obtained using the theoretical formulation described in section I.
The green(thick) line shows the result without MOV contribution and
hence describes the noise correlations for the spin-$1/2$ XY chain.
In this case, $\nu=1/3$ and $2/3$ results coincide reflecting
particle-hole symmetry. } \label{fig4}
\end{figure}

\begin{figure}[htbp]
\begin{center}
\includegraphics[width=3.5 in,height=3.5 in]{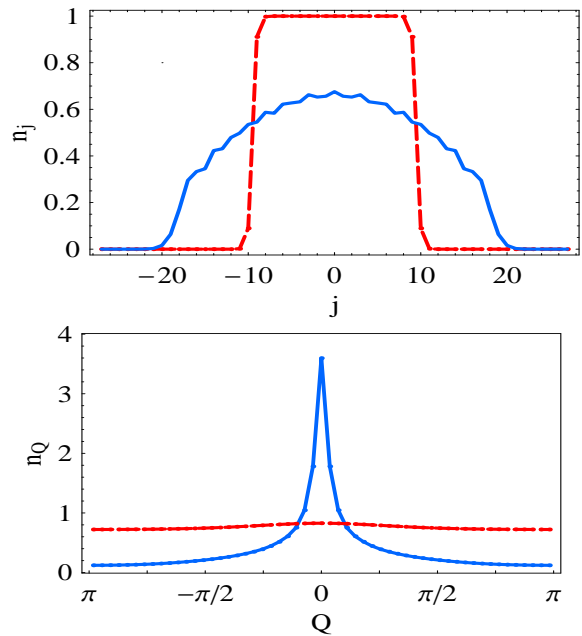}
\leavevmode\end{center} \caption{ Top: Density, Bottom: momentum
distribution (Q is in units of $1/a$) for different trapping
potentials: $0.008$(blue solid line ), and $0.17$ (red
dashed-line)). Here $N=19$ and $L=55$. The correlation functions are
renormalized by a scaling factor $N/Z$ where $Z$ are the number of
sites with non-zero density.} \label{fig5}
\end{figure}

\section{ Mott vs non-Mott phase in a Parabolically confined  Case }
\label{three}
\begin{figure}[htbp]
\begin{center}
\includegraphics[width=4.in]{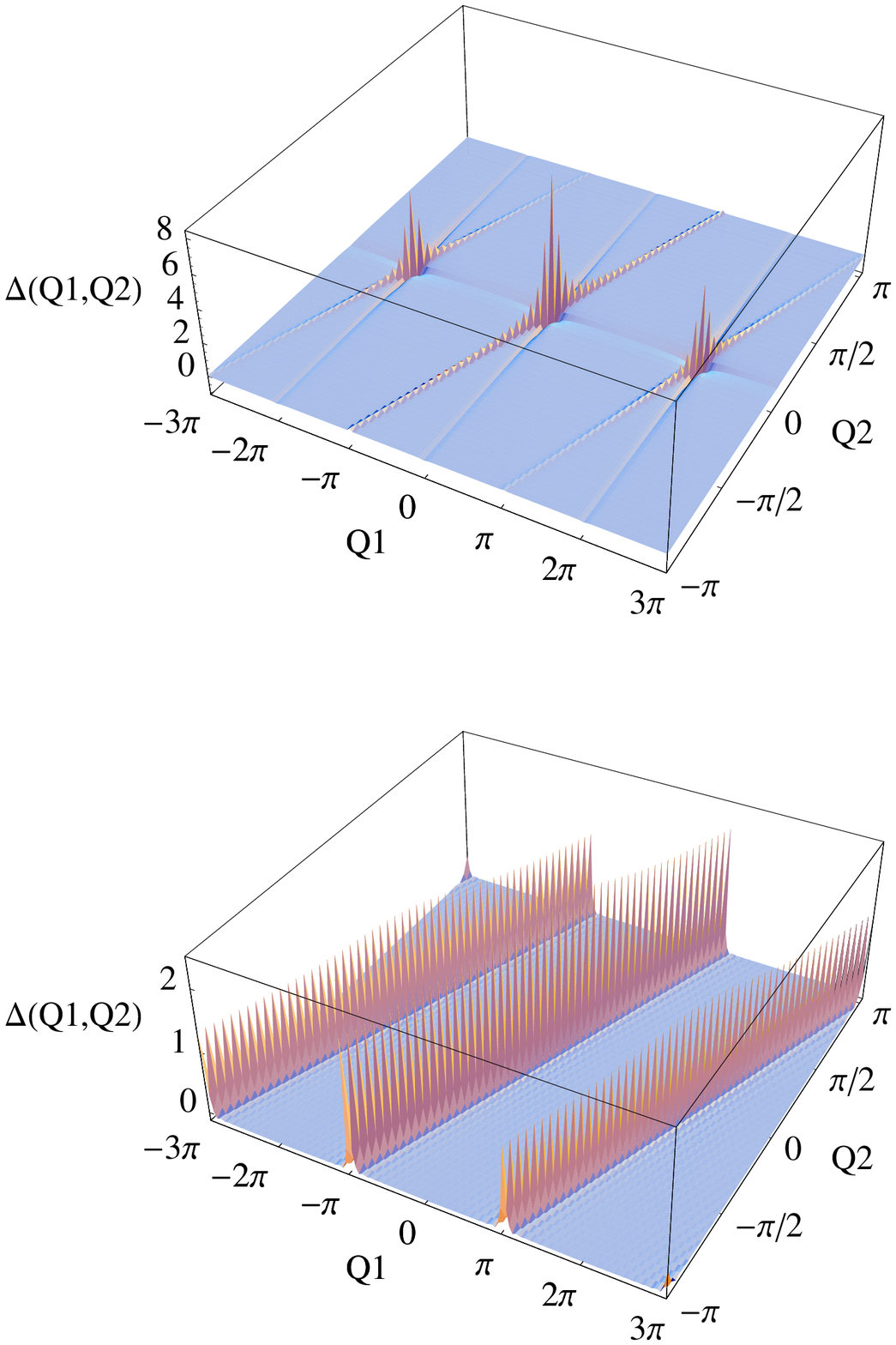}
\leavevmode
\end{center}\caption{  Noise correlations as a function of $Q_1$ and
$Q_2$ for the same  parameters used in Fig.4. $Q_1$ and $Q_2$  are
in units of $1/a$. The  upper and lower correspond to $\Omega/J=
0.008 $( Non MI) and $0.17$(MI) respectively. The correlation
functions are renormalized by a scaling factor $N/Z$ where $Z$ are
the number of sites with non-zero density.} \label{fig6 }
\end{figure}

In this section we study noise correlations for the experimentally
relevant case when there is an additional parabolic confinement
$V_j=\Omega j^2$. The advantage of having the parabolic confinement
is that in this case it is always possible to realize a unit MI at
the trap center for any number of particles with an appropriate
choice of the trapping potentials \cite{Rigol,GAG}. In the
homogeneous case on the contrary only for the integer filled case is
a MI  possible. In our analysis, the chosen parameters correspond to
typical experimental set-ups such  as the ones reported in Ref.
\cite{Paredes}. We present our results for two different values of
$\Omega/J$, 0.008 and 0.17.

The density profiles (Fig. 5 top panel) show that for the case
$\Omega/J=0.17$ the ground state of the system corresponds to a MI
with all the central $N$ sites   singly occupied. For
$\Omega/J=0.008$, on the other hand all, the sites have filling
factor less than unity. The formation of a MI state with reduced
number fluctuations and the localization of the atoms at individual
lattice sites are signaled by the momentum distribution (Fig.5
bottom panel), which shows a flat profile for $\Omega/J=0.17$. On
the other hand, the non-vanishing off diagonal coherence and large
number fluctuations in the $\Omega/J=0.008$ case are reflected by
the sharp peak in the momentum distribution at $Q=0$.

In Fig.6 we plot $\Delta(Q_1,Q_2)$ for the Mott and non-Mott systems
as a function of $Q_1$ and $Q_2$. Noise correlations show
interference peaks when  $Q_1=Q_2 + nK$. The existence  of these
peaks in MI as well as in non-Mott phase implies that the presence
of peaks is not a signature of the insulating phase.
 Our numerical calculations show that, nevertheless, information about
the insulator or superfluid character of the system can be extracted
from noise-interferometry as only when the system is a MI, the
noise-correlations exhibit a regular pattern, i.e.
$\Delta(Q_1,Q_2)\approx \Delta(Q_1-Q_2)$ (small differences seen in
Fig.~6 are due to the finite trap). In contrast, when tunneling is
allowed and  non-local correlations are established, they cause a
decay in the intensity of the peaks as one moves away from
$[Q_1]_K=[Q_2]_K=0$. This decay is similar to  the one observed in
the momentum distribution away from $Q=0$ and reflects the
non-vanishing off-diagonal coherence of the system.

Furthermore,  Fig. 6  shows that the satellite dips observed in
partially filled  translationally invariant systems are still
present in the inhomogeneous case. When $\Omega/J=0.008$ they are
located at $[Q_1]_K=2\pi/(La)$ $[Q_2]_K = 0$ and disappear in the MI
phase.

All of these results suggest that noise interferometry  provides
valuable complementary information about the phase coherence of the
system and thus can be a suitable experimental tool for
characterizing the MI phase. Experimental techniques capable of
thoroughly characterizing the MI phase are crucial for the neutral
atom based quantum information proposals.

\section{Summary}
In this paper, we have derived explicit formulae to calculate
four-point functions in HCB, spin-$1/2$ and fermionic  systems in
the presence of an external potential. Our results are valid for all
filling factors and hence include the Mott as well as the non-Mott
regimes. Our work generalizes the  Lieb and Mattis formulation for
two point correlations to four-point correlations and is therefore
an asset for all future studies of higher order correlations in HCB
as well as in spin-$1/2$ systems.

One of the key results of our studies is the fact that although
multiply occupied states are suppressed in the ground state of
strongly correlated bosons, MOV states have to be included for a
proper calculation of correlations. These states  lead to
differences between HCB and spin-$1/2$ XY chains which have
important manifestations  such as the breaking of particle-hole
symmetry in the HCB systems. Recently, there have been various
proposals to use bosonic atoms in optical lattices as quantum
simulators of spins models \cite{atmspin}. Our study points out that
such  mappings have to be done very carefully as MOV states can
certainly  change the spin character of certain observables.

We also showed that the noise correlation pattern is a sensitive
probe to characterize the MI phase as:  i) only in the MI phase is
the noise-pattern  completely regular, ii) there are additional
satellite dips accompanying the Bragg interference peaks in the
absence of a MI.  These findings might have some impact in current
efforts to implement a quantum computer using optical lattices as
most of proposals require a well characterized MI state to
initialize the quantum register \cite{GAG}.

We would like to emphasize that we have studied experimentally
measurable shot noise correlations for strongly correlated bosons in
the TG regime. Discussions related to the corresponding spin system
is included to highlight the importance of multiple occupancy of
virtual states in HCB, as the correlations in the spin systems
describe correlations of HCB without MOV. In other words, results
for spins are not presented in the context of any experiments in
these systems. However, our general expressions for four-point
correlations for spins will potentially  be very useful for various
studies of higher order correlations in spin systems such as XY and
Ising chains with or without disorder.

\noindent\textbf{Acknowledgments}

We are grateful to Trey Porto for various suggestions and comments.
We would like to thank L. Mathey {\it et. al} \cite{student} for
sharing their results (under preparation) of a related study of
noise correlations using bosonization techniques where  a
characteristic peak, followed by a satellite dip (similar to those
shown in Fig.1) are found. This work is supported in part by the
Advanced Research and Development Activity (ARDA) contract and the
U.S. National Science Foundation through a grant PHY-0100767. A.M.R.
acknowledges additional support by a grant from  the Institute of
Theoretical, Atomic, Molecular and Optical Physics at Harvard
University and Smithsonian Astrophysical observatory.


\begin{thebibliography}{99}
\bibitem{student}
  Similar results using bosonization techniques for 1D systems have
been found by Mathey L,  Altman E  and Vishwanath A . In
preparation.
\end{thebibliography}
\end{document}